\documentclass[12pt,letterpaper]{article}%
\usepackage{amsmath,amssymb,amsfonts}
\usepackage[pdftex]{graphicx}
\usepackage{float}
\usepackage{caption2}
\usepackage[]{hyperref}
\usepackage[Symbolsmallscale]{upgreek}
\usepackage{dsfont}
\usepackage{graphicx}

\setcaptionmargin{1cm}

\begin{document}
\begin{titlepage}
\begin{flushright}
\end{flushright}
\vskip 1.0cm
\begin{center}
{\Large \bf Rigorous Limits on the Interaction Strength\\[0.3cm]in Quantum
Field Theory} \vskip 1.0cm {\large Francesco Caracciolo$^{a,c}$ and Slava Rychkov$^{b,c}$ }\\[0.7cm]
{\it $^a$ SISSA, Trieste, Italy}\\
{\it $^b$
Laboratoire de Physique Th\'{e}orique, Ecole Normale Superieure,\\
and Facult\'{e} de physique, Universit\'{e} Paris VI, France}\\
{\it $^c$ Scuola Normale Superiore and INFN, Sezione di Pisa, Italy}\\
\end{center}
\vskip 2.0cm
\begin{abstract} We derive model-independent, universal upper bounds
on the Operator Product Expansion (OPE) coefficients in unitary
4-dimensional Conformal Field Theories. The method uses the
conformal block decomposition and the crossing symmetry constraint
of the 4-point function. In particular, the OPE coefficient of three
identical dimension $d$ scalar primaries is found to be bounded by
$\simeq 10(d-1)$ for $1<d<1.7$. This puts strong limits on
unparticle self-interaction cross sections at the LHC.
\end{abstract}
\vskip 2cm \hspace{0.7cm} December 2009
\end{titlepage}

\newpage

In this paper we will answer, in a particular well-defined context, the
question:\textit{ Is there an upper bound to the interaction strength in
relativistic Quantum Field Theory (rQFT)?}


Intuitive reasons suggest that such a bound exists. Take QCD as a
representative real-world example. At energies $E$ above the scale
$\Lambda_{\text{QCD}}\sim1$ GeV, this is a perturbative theory of interacting
quarks and gluons, and the interaction strength is measured by the
dimensionless running coupling $g_{s}(E)$. The coupling starts small at very
high energies $E\gg\Lambda_{\text{QCD}}$ and grows at low energies, formally
becoming infinite at $E\sim\Lambda_{\text{QCD}}$. However, perturbative
expansion breaks down before this happens. $L$-loop diagrams are suppressed by
factors $\sim(g_{s}^{2}/16\pi^{2})^{L}$. As soon as $g_{s}\sim4\pi$, all loop
orders contribute equally. Thus in perturbation theory it is impossible to get
couplings stronger than about $4\pi$.

To recall what happens beyond perturbation theory, let us look at the same
theory at energies below $\Lambda_{\text{QCD}}$. In this regime the
appropriate degrees of freedom are hadrons, and their interactions can be
described by an effective lagrangian. For instance, pion-pion scattering at
low energies is described by the chiral lagrangian%
\[
\mathcal{L}=\frac{f_{\pi}^{2}}{4}\text{Tr}|\partial_{\mu}U|^{2}+\ldots,\quad
U=\exp(i\,2\,\pi^{a}T^{a}/f_{\pi}),
\]
where $f_{\pi}\simeq93$ MeV is the pion decay constant, $T^{a}$ are the
\textit{SU(2)} generators and \ldots\ stand for the chiral symmetry breaking
terms. The dimensionless quartic pion coupling can be defined from the
$2\rightarrow2$ scattering amplitude; it grows with energy as $\lambda
\sim(E/f_{\pi})^{2}.$ If the chiral lagrangian is valid up to energies
$\sim\Lambda_{\text{QCD}}$ and is stable under radiative corrections, we
should have $\lambda(\Lambda_{\text{QCD}})/16\pi^{2}\lesssim1$, or
$\Lambda_{\text{QCD}}\lesssim4\pi f_{\pi}$. Experimentally this bound is
satisfied and, moreover, near-saturated. This observation forms the basis of
the Naive Dimensional Analysis \cite{NDA} method of estimating couplings in
strongly coupled theories.

While the above arguments are appealing, at present it is unknown if they can
be turned into a theorem, or even how to formulate such a general theorem. In
order to make progress, in what follows we will assume that we have a
Conformal Field Theory (CFT), i.e. an rQFT invariant under the action of the
conformal group \cite{df}.

CFTs form an important subclass of rQFTs. Presumably, any unitary, scale
invariant rQFT is conformally invariant. This is proved in $D=2$ spacetime
dimensions under very mild technical assumptions \cite{Zamolodchikov:1986gt}%
,\cite{Polchinski:1987dy}, and no counterexamples are known in $D\geq3.$ Since
scale invariance\ is ubiquitous (think of any RG-flow fixed point), this would
make conformal invariance equally ubiquitous. Unitarity is however crucial
here: without unitarity simple physical counterexamples exist, e.g.~theory of
elasticity \cite{Riva:2005gd}. We are interested in applications to particle
physics, thus we will assume unitarity, and will work in $D=4$.

There are many known or conjectured classes of four-dimensional CFTs. For
example, $\mathcal{N}=1$ supersymmetric QCD with $N_{c}$ colors and $N_{f}$
flavors flow to a CFT in the infrared as long as $3/2<N_{f}/N_{c}<3$
\cite{seiberg}. Large $N_{c}$ analysis \cite{belavin} and lattice simulations
\cite{lattice} suggest that a similar `conformal window' exists also without
supersymmetry. Another famous example is the $\mathcal{N}=4$ super Yang-Mills
(SYM), which is conformal for any coupling and any $N_{c}$. At large 't Hooft
coupling and large $N_{c}$ it can be described via the AdS/CFT correspondence.
Many deformations preserving conformal symmetry are known on both field theory
and gravity sides of the correspondence \cite{adscft}. Our discussion will be
general and will in principle apply to all the above examples.

The $D=4$ conformal group is finite dimensional; it is obtained from the
Poincar\'{e} group by adding the generators of dilatation $\mathcal{D}$ and of
special conformal transformations $\mathcal{K}_{\mu}$. The local quantum
fields $O(x)$ are eigenstates of $\mathcal{D}$, $[\mathcal{D},O(0)]=i\Delta
O(0)$, where the eigenvalue $\Delta$ is called the scaling dimension. The
$\mathcal{K}_{\mu}$ acts as a lowering operator for the scaling dimension, and
the corresponding `lowest-weight states'\textit{,} fields satisfying
$[\mathcal{K}_{\mu},O(0)]=0$\textit{,} play a special role. They are called
\textit{primaries}. All other fields can be obtained from primaries by taking
derivatives and are called \textit{descendants}.

Conformal symmetry constraints the 2- and 3-point functions of primary fields
to have particularly simple form. For scalar primaries, we have:%
\begin{gather}
\left\langle O_{i}(x_{1})O_{j}(x_{2})\right\rangle =\delta_{ij}(x_{12}%
^{2})^{-\Delta}~,\label{eq:2pt}\\
\left\langle O_{i}(x_{1})O_{j}(x_{2})O_{k}(x_{3})\right\rangle =c_{ijk}%
\,(x_{12}^{2})^{\rho_{kij}}(x_{13}^{2})^{\rho_{jik}}(x_{23}^{2})^{\rho_{ijk}%
}~,\label{eq:3pt}\\
x_{ij}^{2}\equiv(x_{i}-x_{j})^{2},\qquad\rho_{ijk}\equiv(\Delta_{i}-\Delta
_{j}-\Delta_{k})/2\,.\nonumber
\end{gather}
Eq.~(\ref{eq:2pt}) says that a diagonal basis can be chosen in the space of
primary fields, and sets the normalization. Eq.~(\ref{eq:3pt}) then defines
coefficients $c_{ijk}$. These same coefficients appear in the Operator Product
Expansion (OPE)%
\[
O_{i}(x)O_{j}(0)\sim(x^{2})^{-(\Delta_{i}+\Delta_{j})/2}\left\{
{\mathds1}+c_{ijk}(x^{2})^{\Delta_{k}/2}O_{k}(0)\,+\ldots\right\}  \text{,}%
\]
where \ldots\ stands for the contributions of higher spin primaries and of descendants.

In CFT, any $n$-point function can be, in principle, reduced to a sum of
products of 2-point functions by repeated application of the OPE, with
coefficients given by products of $c_{ijk}$'s and of their higher spin
generalizations. In this sense, $c_{ijk}$'s play in CFT a role similar to that
of the (renormalized) coupling constants in perturbation theory, measuring
interaction strength. We thus have the following CFT version of our initial
question: \textit{Is there an upper bound to the OPE coefficients, valid in an
arbitrary unitary CFT in }$D=4$\textit{? }We will now proceed to show that
such a universal bound indeed exists.

Let us pick a hermitean scalar primary $\phi$ of scaling dimension $d$ and
consider its OPE with itself:%
\begin{equation}
\phi(x)\phi(0)\sim(x^{2})^{-d}\Bigl\{{\mathds1}+\sum_{l=0,2,4\ldots}%
\sum_{\Delta\geq\Delta_{\min}(l)}c_{\Delta,l}(x^{2})^{(\Delta-l)/2}x^{\mu_{1}%
}\cdots x^{\mu_{l}}O_{\mu_{1}\ldots\mu_{l}}(0)\,+\ldots\Bigr\}\,.
\label{eq:phiphi}%
\end{equation}
This time we show explicitly contributions of both scalars $(l=0)$ and of
higher spin primaries $O_{\mu_{1}\ldots\mu_{l}}$ which are symmetric traceless
tensors. Spin $l$ has to be even by the Bose symmetry. Lower bounds on the
dimension $\Delta$ of a spin $l$ primary:%
\[
\Delta_{\min}(l=0)=1,\quad\Delta_{\min}(l\geq1)=l+2\,\text{,}%
\]
(\textit{unitarity bounds) }are known to follow from unitarity \cite{mack}.
Only special fields may saturate these bounds: a free scalar $(l=0)$,
conserved currents $(l=1)$, and the stress tensor $(l=2).$ Higher $l$
conserved currents, present in free theories, also saturate the bounds.

An interesting object to study is the 4-point function of $\phi$, constrained
by conformal symmetry to have the form%
\begin{equation}
\left\langle \phi(x_{1})\phi(x_{2})\phi(x_{3})\phi(x_{4})\right\rangle
=\frac{g(u,v)}{x_{12}^{2d}x_{34}^{2d}}, \label{eq:4pt}%
\end{equation}
where $u=x_{12}^{2}x_{34}^{2}/(x_{13}^{2}x_{24}^{2}),$ $v=x_{14}^{2}x_{23}%
^{2}/(x_{13}^{2}x_{24}^{2})$ are the conformal cross-ratios$.$ The same
4-point function can be reduced to a sum of 2-point functions by applying the
OPE in the 12 and 34 channels. Cross terms of different primary families drop
out of this representation because of Eq.~(\ref{eq:2pt}) and its higher spin
analogue. Terms involving the same primary and its descendants can be resummed
in closed form. As a result, we get the \textit{conformal block decomposition}%
\begin{equation}
g(u,v)=1+\sum p_{\Delta,l}\,g_{\Delta,l}(u,v)\,,\quad p_{\Delta,l}\equiv
c_{\Delta,l}^{2}\,, \label{eq:cb}%
\end{equation}
where \cite{do1}%
\begin{gather*}
g_{\Delta,l}(u,v)=\frac{(-)^{l}}{2^{l}}\frac{z\bar{z}}{z-\bar{z}}\left[
\,k_{\Delta+l}(z)k_{\Delta-l-2}(\bar{z})-(z\leftrightarrow\bar{z})\right]
\,,\\
k_{\beta}(x)\equiv x^{\beta/2}{}_{2}F_{1}\left(  \beta/2,\beta/2,\beta
;x\right)  \,,\\
u=z\bar{z},\quad v=(1-z)(1-\bar{z})\,.
\end{gather*}
This decomposition is expected to converge at least in the circle
$|z|<1$,$|\bar{z}|<1$, which corresponds to being able to fit a sphere
centered at $x_{1}$ which separates $x_{2}$ from $x_{3}$ and $x_{4}$
\cite{pol1}.

The 4-point function (\ref{eq:4pt}) must be crossing-symmetric under the
$x_{1}\leftrightarrow x_{2}$ and $x_{1}$ $\leftrightarrow x_{3}$ exchanges.
The first crossing is manifest since only even spins contribute to the OPE.
The second one gives a nontrivial constraint%
\begin{equation}
v^{d}g(u,v)=u^{d}g(v,u)\,. \label{eq-boot}%
\end{equation}
Decomposition (\ref{eq:cb}) must be consistent with this constraint.
Separating the contribution of the unit operator, we obtain the \textit{sum
rule}
\begin{align}
1  &  =\sum p_{\Delta,l}F_{d,\Delta,l}(u,v)\,,\quad\label{eq:sum}\\
F_{d,\Delta,l}(u,v)  &  \equiv\frac{v^{d}g_{\Delta,l}(u,v)-u^{d}g_{\Delta
,l}(v,u)\,}{u^{d}-v^{d}}. \label{eq-F}%
\end{align}
As we will now show, this equation can be used to get an upper bound on
$c_{\Delta,l}$.

Crucially, coefficients $c_{\Delta,l}$ are real, and thus $p_{\Delta,l}\geq0$.
This can be related to the absence of parity violation in the conformal
3-point function of two scalars and a symmetric tensor \cite{r1}.
Eq.~(\ref{eq:sum}) then allows a geometric interpretation: when $p_{\Delta
,l}\geq0$ are allowed to vary, the RHS fills a convex cone $C_{d}$ in the
vector space $\mathcal{V}$ whose elements are two-variable functions. We say
that this cone is \textit{generated} by functions $F_{d,\Delta,l}(u,v)$,
$l=0,2,4,\ldots,\Delta\geq\Delta_{\min}(l)$. Eq.~(\ref{eq:sum}) expresses the
fact that the function $f(u,v)\equiv1$ belongs to this cone.

Let us now pick a particular field $O_{\bar{\Delta},\bar{l}}$ and rewrite
(\ref{eq:sum}) as%
\begin{equation}
1-p_{\bar{\Delta},\bar{l}}F_{d,\bar{\Delta},\bar{l}}(u,v)=\sum p_{\Delta
,l}F_{d,\Delta,l}(u,v)\,. \label{eq:sum1}%
\end{equation}
As $p_{\bar{\Delta},\bar{l}}$ is increased, the vector corresponding to
$1-p_{\bar{\Delta},\bar{l}}F_{d,\bar{\Delta},\bar{l}}(u,v)$ moves in the
vector space. Suppose that for all $p_{\bar{\Delta},\bar{l}}$ above some
critical value $p_{\text{cr}}$ this vector stays out of the cone $C_{d}$. Then
$p_{\text{cr}}$ provides a bound on the squared OPE coefficient $|c_{\bar
{\Delta},\bar{l}}$%
$\vert$%
$^{2}$. This bound will depend on $d,\bar{\Delta},\bar{l}$, but will be valid
in any unitary CFT.

To find $p_{\text{cr}}$, we employ the method of linear functionals developed
in \cite{r1},\cite{r2}. Recall that a linear functional is a linear map
$\Lambda$ from $\mathcal{V}$ to real numbers:%
\begin{equation}
\Lambda:\mathcal{V}\rightarrow\mathbb{R},\quad\Lambda\lbrack\alpha_{i}%
F_{i}]=\alpha_{i}\Lambda\lbrack F_{i}]\,.
\end{equation}
Suppose that we found a functional which is positive on all functions
generating the cone $C_{d}$:%
\begin{equation}
\Lambda\lbrack F_{d,\Delta,l}]\geq0\,. \label{eq:pos}%
\end{equation}
We will normalize this functional by the condition%
\begin{equation}
\Lambda\lbrack1]=1. \label{eq:norm}%
\end{equation}
Since for such $\Lambda$ Eq.~(\ref{eq:sum}) implies $\Lambda\lbrack
1-p_{\bar{\Delta},\bar{l}}F_{d,\bar{\Delta},\bar{l}}]\geq0$, we would get an
upper bound:%
\begin{equation}
p_{\bar{\Delta},\bar{l}}\leq p_{\text{cr}}(\Lambda)\equiv1/\Lambda\lbrack
F_{d,\bar{\Delta},\bar{l}}]. \label{eq:bound}%
\end{equation}
To make this bound as strong as possible, we will impose, in addition to
(\ref{eq:pos}), (\ref{eq:norm}), an extremality condition%
\begin{equation}
\Lambda\lbrack F_{d,\bar{\Delta},\bar{l}}]\rightarrow\max\,. \label{eq:max}%
\end{equation}

We will use linear functionals given by a finite linear combination of
derivatives evaluated at a given point. More precisely, we will use
functionals of the form:%
\begin{gather}
\Lambda\lbrack F]\equiv\sum_{n,m\geq\,0,~n+m\leq\,N}\lambda_{n,m}%
F^{(2n,2m)},\quad N=3\,,\label{eq:funct}\\
F^{(2n,2m)}\equiv\partial_{a}^{2n}\partial_{b}^{2m}F|_{a=b=0},\\
z=1/2+a+b,\quad\bar{z}=1/2+a-b\,.
\end{gather}
Here $\lambda_{n,m}$ are fixed real numbers defining the functional. The
symmetric point $a=b=0$ is chosen as in \cite{r1},\cite{r2} since the sum rule
is expected to converge fastest here, and because the functions $F_{d,\Delta
,l}$ are even in both variables with respect to this point. This is why only
even-order derivatives are included in (\ref{eq:funct}).

Eqs.~(\ref{eq:pos}), (\ref{eq:norm}), (\ref{eq:max}) define an optimization
problem for the coefficients $\lambda_{n,m}$. The constraints are given by
linear equations and inequalities, and the cost function is also linear, which
makes it a linear programming problem. Although the number of constraints in
(\ref{eq:pos}) is formally infinite, they can be reduced to a finite number by
discretizing $\Delta$ and truncating at large $\Delta$ and $l$, where the
constraints approach a calculable asymptotic form. The reduced problem can be
efficiently solved by well-known numerical methods, such as the simplex
method. A found solution can be then checked to see if it also solves the full
problem. This procedure was developed and successfully used in a different
context in \cite{r1},\cite{r2}.

In this work, we used this procedure to compute bounds on the OPE coefficients
$c_{\phi\phi O}$ when $O$ is a scalar field $(\bar{l}=0$). We will now present
our numerical results. Fig.~1 concerns the case when the dimension of $\phi$
is close to that of a free field, $1<d\leq1.1$. Notice the bell-shaped form of
the bound, peaked at $\bar{\Delta}\simeq2$.\footnote{This shape makes it
tempting to draw an analogy with the Breit-Wigner formula, especially since
the dilatation operator $\mathcal{D}$ plays the role of energy in radial
quantization.} For $d\rightarrow1$ the bound evidently tends to zero
everywhere except near $\bar{\Delta}=2$. This means that the free field theory
limit is approached continuously: for $d=1$ the only scalar operator in the
$\phi\times\phi$ OPE is the $:\!\phi^{2}\!:$ of dimension $2$. In Fig.~2 we
present a similar plot for $1.2\leq d\leq1.7$. Notice that the bounds in
Figs.~1,2 go to zero as $\bar{\Delta}\rightarrow1$. This is expected in view
of the general theorem that a dimension 1 scalar must be free, hence decoupled
from everything else in the CFT.

\begin{figure}[ptb]
\begin{center}
\includegraphics[width=0.55 \textwidth]{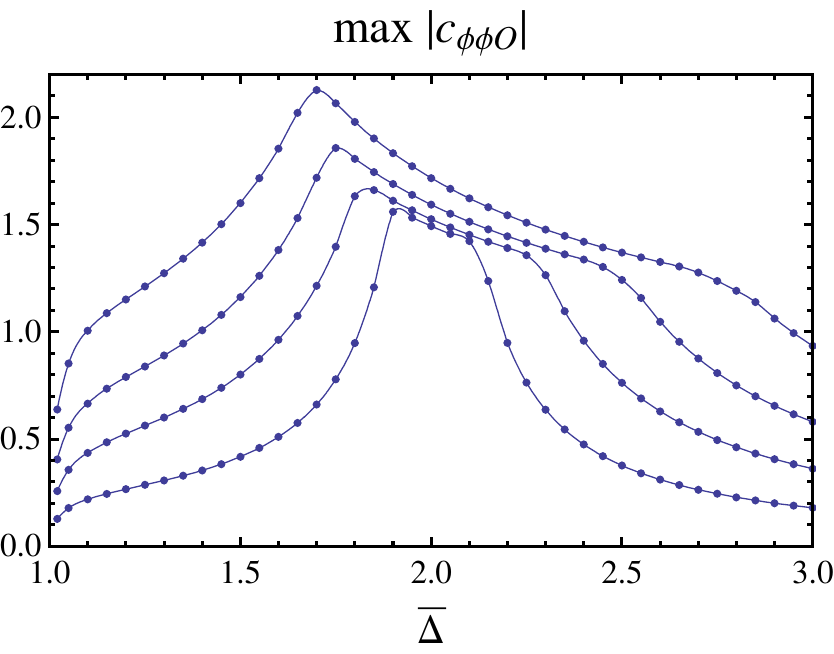}
\end{center}
\caption{\textit{Theoretical upper bound for the OPE coefficient $c_{\phi\phi
O}$ as a function of the dimension $\bar{\Delta}$ of the scalar field }%
$O$\textit{. The curves correspond to the $\phi$'s dimension fixed at
$d=1.005,\,1.02,\,1.05,\,1.1$ (from below up). The bound was computed for each
of the shown points, and the curves in between were obtained by interpolation.
}}%
\end{figure}

\begin{figure}[ptb]
\begin{center}
\includegraphics[width=0.55 \textwidth]{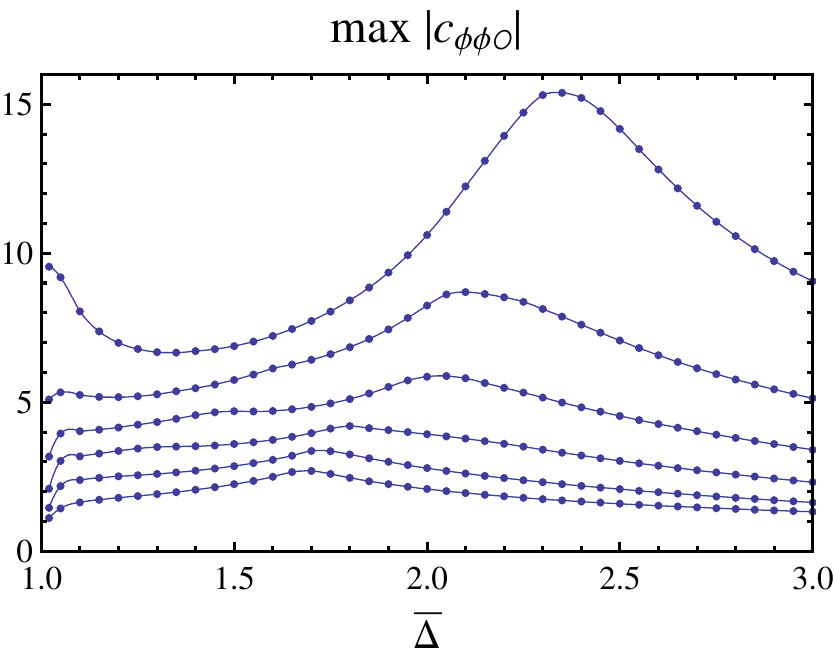}
\end{center}
\caption{\textit{Same as Fig.\ 1 for the $\phi$'s dimension fixed at
$d=1.2\,,1.3\,,1.4\,,1.5\,,1.6\,,1.7$ (from below up). }}%
\end{figure}

\begin{figure}[ptb]
\begin{center}
\includegraphics[width=0.53 \textwidth]{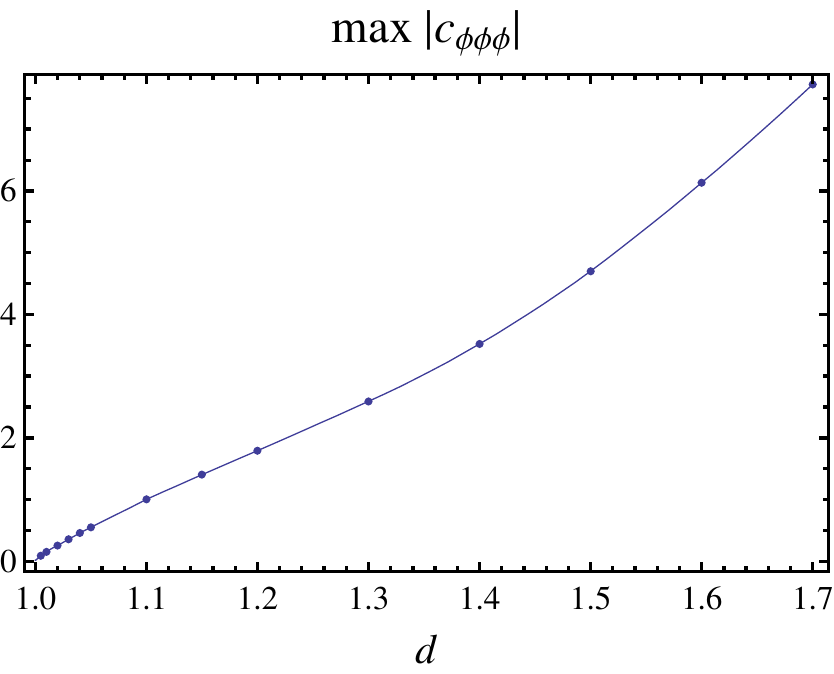}
\end{center}
\caption{\textit{Theoretical upper bound for the OPE coefficient $c_{\phi
\phi\phi}$ as a function of $\phi$'s dimension $d$.}}%
\end{figure}

A text file with the coefficients of the linear functionals used to obtained
the bounds plotted in Figs.~1,2 is included in the source file of this arXiv
submission. The reader may check that they indeed satisfy the constraints
(\ref{eq:pos}).

We have only explored the range $d\leq1.7$ for the following reason: starting
from $d\simeq1.75$, we found that there is no functional of the form
(\ref{eq:funct}) satisfying the constraints (\ref{eq:pos}), (\ref{eq:norm}).
We expect that a bound exists also for larger $d$, but to find it one needs to
use more general functionals, e.g.~involving more derivatives (i.e.~with
higher $N$). This will also give improved bounds in the range of $d$ that we
considered$.$ This is left for future work.

On the other hand, the restriction to $1\leq\bar{\Delta}\leq3$ in Figs.~1,2 is
not essential: our method would also give bounds beyond this range. In fact,
any of the functionals derived for $1\leq\bar{\Delta}\leq3$ could be used to
compute a\textit{ }sub-optimal but valid bound for larger $\bar{\Delta}$ (as
well as for $\bar{l}>0$) via Eq.~(\ref{eq:bound}).

It would be interesting to study the asymptotic behavior of the bound at large
$\bar{\Delta}$. A conservative upper estimate can be obtained from the known
asymptotics of $F_{d,\bar{\Delta},\bar{l}}$ and its derivatives \cite{r1}, if
we assume that the functional $\Lambda$ in (\ref{eq:bound}) is $\bar{\Delta}%
$-independent. This way one concludes that the bound cannot grow faster than
exponentially: $|c_{\phi\phi O}|=\mathcal{O}(q^{\bar{\Delta}}),$ $q=(\sqrt
{2}+1)/2$. However, this is likely an overestimate, since the optimal
functional $\Lambda$, as determined by Eq.~(\ref{eq:max}), will likely depend
on $\Lambda$.

It would be also interesting to derive analogous bounds in two spacetime
dimensions, where explicit expressions for conformal blocks are also known
\cite{do1}.

As a phenomenoligical application of our results, consider the unparticle
physics scenario \cite{me-too-physics}. Unparticle self-interactions were
considered in \cite{Feng} (see also \cite{self-int-th}) a prominent feature of
such scenarios, giving rise to processes like $gg\rightarrow\phi
\rightarrow\phi\phi\rightarrow4\gamma$. The cross section for this process is
proportional to the square of the self-coupling OPE coefficient $c_{\phi
\phi\phi}$, where $\phi$ is a scalar operator from a hidden-sector CFT
(\textit{unparticle}) with non-renormalizible couplings to gluons and photons.
In \cite{Feng}, the values of these coefficients were kept as arbitrary
parameters, unconstrained by prime principles, and only experimental
constraints from the Tevatron were imposed, which led to a possibility of
spectacularly large cross sections at the LHC. In Fig. 3 we plot our
theoretical upper bound on $c_{\phi\phi\phi}$ (extracted from Figs.~1,2 by
setting $\bar{\Delta}=d$). The values of $c_{\phi\phi\phi}$ used in
\cite{Feng} exceed our bound by $2\div4$ orders of magnitude.\footnote{For
proper comparison note that the normalization of the unparticle OPE
coefficients $C_{d}$ used in \cite{Feng} is related to our normalization via
$C_{d}=g_{d/2}^{3}(|B_{d}|/g_{d})^{3/2}c_{\phi\phi\phi}$ where $B_{d}$ is
given in \cite{Feng} and $g_{d}=4^{2-d}\pi^{2}\Gamma(2-d)/\Gamma(d)$.} We
conclude that a revision of the studies in \cite{Feng},\cite{self-int-ph},
taking into account our bounds, is necessary.

As a purely field-theoretical application, consider the $\mathcal{N}=4$ SYM
theory already mentioned above, which is conformal for any value of the 't
Hooft coupling $\lambda=g_{YM}^{2}N_{c}$. The region of small $\lambda$ is
accessible via perturbation theory, while large $\lambda$ (and large $N_{c})$
are accessible via the AdS/CFT correspondence. Moreover, the large $N_{c}$
theory is integrable, which allows to interpolate between the two regimes and
perform various nontrivial checks \cite{kaz}. As $\lambda$ is increased from
$0$ to $\infty$, the spectrum of the theory changes, and anomalous dimensions
of some local fields are certain to become large. For example, at large
$N_{c}$ the fields which do not map onto supergravity modes on $AdS_{5}\times
S^{5}$ have anomalous dimensions growing for large $\lambda$ as $\lambda
^{1/4}$ \cite{GKP}. Now one could ask what happens to the OPE coefficients,
whether they can have similar growth. From our results, assuming that they can
be extended to $d>1.7$ as discussed above, it follows that no matter how large
$\lambda$ is, the OPE coefficients \textit{of fields with low dimensions} will
stay bounded. It should be noted that this conclusion is nontrivial only for
small $N_{c}$, since at large $N_{c}$ the OPE $O_{1}\times O_{2}$ is known to
factorize, with the composite \textquotedblleft multi-trace" fields
$:O_{1}O_{2}:$ appearing with the coefficient $1+\mathcal{O}(1/N_{c}^{2}$)
while all other fields $1/N_{c}$ suppressed \cite{fact}.

In summary, we have presented theoretical upper bounds on the OPE coefficients
of two identical scalars and a third scalar, valid in an arbitrary unitary
CFT. Our results are based on imposing crossing symmetry on the conformal
block decomposition of a scalar 4-point function. They imply that, in a
certain sense, interaction strength remains limited even in theories like
$\mathcal{N}=4$ SYM (or its many known conformal deformations) where a
coupling $\lambda$ can be taken to infinity. They also lead to strong bounds
on the cross sections of unparticle self-interaction-type processes at future colliders.

\subsection*{Acknowledgements}

We~are grateful to R.~Rattazzi, A.~Vichi and K.~Zarembo for useful discussions
and comments on the draft, and to R.~Rattazzi for suggesting the Breit-Wigner
analogy. This research was supported in part by the European Programme
"Unification in the LHC Era", contract PITN-GA-2009-237920 (UNILHC).

\end{document}